\documentclass[aps,reprint,floatfix,superscriptaddress,noshowpacs,prl]{revtex4-1}
\usepackage{amsmath, amssymb}
\usepackage{graphicx}
\newcommand*{\ybco}[1]{YBa$_2$Cu$_3$O$_{#1}$}

\begin{document} 

\preprint{ver.~12}

\title{\boldmath Discovery of slow magnetic fluctuations and critical slowing down\\ in the pseudogap phase of \ybco{y}}

\author{Jian Zhang}
\author{Z. F. Ding}
\author{C. Tan}
\author{K. Huang}
\affiliation{State Key Laboratory of Surface Physics, Department of Physics, Fudan University, Shanghai 200433, People's Republic of China}
\author{O. O. Bernal}
\affiliation{Department of Physics and Astronomy, California State University, Los Angeles, California 90032, USA}
\author{P.-C. Ho}
\affiliation{Department of Physics, California State University, Fresno, California 93740,USA}
\author{G. D. Morris}
\affiliation{TRIUMF, Vancouver, BC V6T 2A3, Canada}
\author{A. D. Hillier}
\author{P. K. Biswas}
\author{S. P. Cottrell}
\affiliation{ISIS Facility, STFC Rutherford Appleton Laboratory, Harwell Science and Innovation Campus, Chilton, Didcot, Oxon., UK}
\author{H. Xiang}
\affiliation{State Key Lab for Metal Matrix Composites, Key Lab of Artificial Structures $\&$ Quantum Control (Ministry of Education), Dept.\ of Physics and Astronomy, Shanghai Jiao Tong University, Shanghai 200240, People's Republic of China}
\author{X. Yao}
\affiliation{State Key Lab for Metal Matrix Composites, Key Lab of Artificial Structures $\&$ Quantum Control (Ministry of Education), Dept.\ of Physics and Astronomy, Shanghai Jiao Tong University, Shanghai 200240, People's Republic of China}
\affiliation{Collaborative Innovation Center of Advanced Microstructures, Nanjing 210093, People's Republic of China}
\author{D. E. MacLaughlin}
\email{macl@physics.ucr.edu}
\affiliation{Department of Physics and Astronomy, University of California, Riverside, California 92521, USA}
\author{Lei Shu}
\email{leishu@fudan.edu.cn}
\affiliation{State Key Laboratory of Surface Physics, Department of Physics, Fudan University, Shanghai 200433, People's Republic of China}
\affiliation{Collaborative Innovation Center of Advanced Microstructures, Nanjing 210093, People's Republic of China}
\date{\today}

\begin{abstract}
Evidence for intra-unit-cell (IUC) magnetic order in the pseudogap region of high-$T_c$ cuprates below a temperature~$T^\ast$ is found in several studies, but NMR and $\mu$SR experiments do not observe the expected static local magnetic fields. It has been noted, however, that such fields could be averaged by fluctuations. Our measurements of muon spin relaxation rates in single crystals of YBa$_2$Cu$_3$O$_y$ reveal magnetic fluctuations of the expected order of magnitude that exhibit critical slowing down at $T^\ast$. These results are strong evidence for fluctuating IUC magnetic order in the pseudogap phase.
\end{abstract}

\maketitle

The origin of the pseudogap region below a temperature $T^\ast$ is at the heart of the mysteries of high-$T_c$ cuprate superconductors~\cite{Keimer15}. The predicted broken time-reversal and inversion symmetry due to ordered loop currents~\cite{[{See, e.g., }] [{ and references therein.}]Varma14} or other similar intra-unit-cell (IUC) magnetic order~\cite{Moskvin12, *Fechner16} is consistent with five different classes of symmetry-sensitive experiments: polarized neutron diffraction~\cite{Fauque06, *M-TSWB15}, optical birefringence~\cite{Lubashevsky14}, dichroic ARPES~\cite{Kaminski02}, second harmonic generation~\cite{Zhao16}, and polar Kerr effect~\cite{Xia08}. On the other hand, muon spin rotation ($\mu$SR)~\cite{MacDougall08, *[{}] [{ and references therein.}] PAPI16} and NMR~\cite{Mounce13,WMKH15} experiments do not see the static local fields expected for magnetic order, leaving room for skepticism~\cite{Keimer15}.

A possible cause of the absent static field in the magnetic resonance experiments is fluctuation among alternate orientations of the IUC magnetic order, which averages the local field~$B_\mathrm{loc}(t)$ to zero~\cite{Mounce13}. Such fluctuations can be due to finite-size domains of an ordered phase with different field orientations, as seen in tunneling spectroscopy~\cite{Davis}. For NMR and $\mu$SR, the time scale for such averaging is considerably longer (1--10~$\mu$s) than for the other techniques (0.01--0.1~ns). All experiments would be consistent if the fluctuations were ``slow'', with a characteristic correlation time~$\tau_c$ between these limits. 

Averaging or ``motional narrowing'' of the local field occurs in the rapid-fluctuation limit~$\gamma B_\mathrm{loc}^\mathrm{\,rms}\tau_c \ll 1$, where $\gamma$ is the gyromagnetic ratio of the nucleus or muon and $B_\mathrm{loc}^\mathrm{\,rms} = \langle B_\mathrm{loc}^{\,2}(t)\rangle^{1/2}$ is the rms local field. For the muon $\gamma = \gamma_\mu = 8.5156\times 10^8~\mathrm{s^{-1}~T^{-1}}$. The ordered moments per triangular plaquette obtained from neutron diffraction, 0.05--0.1$\mu_B$ in \ybco{6.66}~\cite{Fauque06} and progressively lower at higher $y$, give dipolar field values~$B_\mathrm{loc} = 1$--10~mT at candidate muon sites in the unit cell~\cite{MacDougall08}. For such fields the above inequality is satisfied for $\tau_c \lesssim 10^{-6}$~s, which is in the range of experimental consistency. 

Then $B_\mathrm{loc}(t)$ gives rise to dynamic or ``spin-lattice'' nuclear or muon spin relaxation. Itoh \textit{et~al.}~\cite{IMY17u} have reported ``ultra-slow'' fluctuations in HgBa$_{2}$CaCu$_{2}$O$_{6+\delta}$ that may be of this kind. In $\mu$SR experiments the motionally-narrowed dynamic muon spin relaxation rate in zero applied field (ZF) is given by $\lambda_\mathrm{ZF} = \gamma_\mu^2 \langle B_\mathrm{loc}^{\,2}\rangle\tau_c$~\cite{Slichter96}. 

The relaxation rate measured in a magnetic field parallel to the initial muon polarization [longitudinal field (LF)] depends on the magnitude~$H_L$ of the field, an effect of sweeping the muon Zeeman frequency through the fluctuation noise power spectrum~\cite{Hayano79,Slichter96}. For Markovian fluctuations with a single well-defined correlation time~$\tau_c$, the LF relaxation rate~$\lambda_\mathrm{LF}$ in a field~$H_L$ is given by the so-called Redfield relation~\cite{Slichter96}
\begin{equation} \label{eq:H}
 \lambda_\mathrm{LF}(H_L) = \frac{\gamma_\mu^2 \langle B_\mathrm{loc}^{\,2}\rangle\tau_c}{1 +(\gamma_{\mu}H_L\tau_c)^2}\,.
\end{equation}
The dependence of $\lambda_\mathrm{LF}$ on $H_L$, if observed to be of the form of Eq.~(\ref{eq:H}), provides estimates of $\tau_c$ and $\langle B_\mathrm{loc}^{\,2}\rangle$. More generally one expects a decrease of $\lambda_\mathrm{LF}$ for $\gamma_\mu H_L$ greater than a characteristic fluctuation rate~$1/\tau_c$, in which case $\tau_c$ and $B_\mathrm{loc}^\mathrm{\,rms}$ from fits of Eq.~(1) to the data are heuristic estimates of the characteristic time and field scales.

This Letter reports the discovery of slow fluctuations via $\mu$SR measurements of $\lambda_\mathrm{LF}(H_L)$ and $\lambda_\mathrm{ZF}$ in single crystals of \ybco{y}, $y = 6.72$, 6.77, 6.83, and 6.95 (superconducting transition temperatures~$T_c = 73$~K, 80~K, 88~K, and 91~K, respectively). Consistency with Eq.~(\ref{eq:H}) is found, and $\tau_c$ and $B_\mathrm{loc}^\mathrm{\,rms}$ are obtained. We also find maxima at $T_\mathrm{mag} \approx T^\ast$ in the temperature dependences of the rates. This is consistent with critical slowing down of magnetic fluctuations near the transition, and demonstrates that these fluctuations are associated with the IUC order.

LF-$\mu$SR experiments were performed on samples with $y = 6.72$, 6.77, and 6.83. The field dependence of $\lambda_\mathrm{LF}$ was measured for these samples at temperatures below $T^\mathrm\ast$ and above $T_c$ for $2~\text{mT} \leqslant \mu_0H_L \leqslant 400$~mT\@. The minimum field was chosen to be much larger than the $\sim$0.1~mT quasistatic field from nuclear dipoles (see Supplemental Material), so that the dipolar field is completely ``decoupled''~\cite{Hayano79}, i.e., the resultant field is nearly parallel to the muon spin and the dipolar fields do not cause appreciable muon precession. Then one observes only dynamic relaxation.

\begin{figure*} [ht]
 \begin{center}
 \includegraphics[clip=,width=0.9\textwidth]{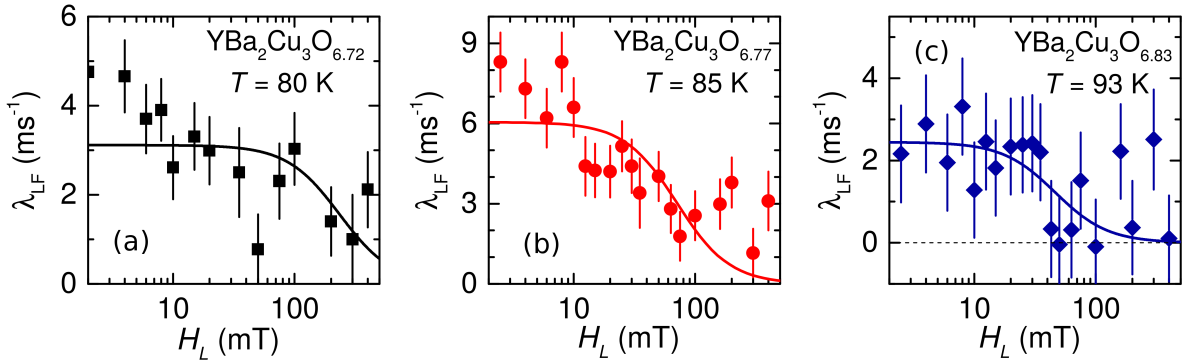}
\caption{\label{fig:LF} (Color online) Dependence of the LF exponential relaxation rate~$\lambda_\mathrm{LF}(H_L)$ on longitudinal magnetic field~$H_L$ in YBa$_2$Cu$_3$O$_y$. (a)~$y=6.72$, $T= 80$~K\@. (b)~$y=6.77$, $T= 85$~K\@. (c)~$y=6.83$, $T= 93$~K\@. Curves: fits of Eq.~(\ref{eq:H}) to the data.}
 \end{center}
\end{figure*}

The results are shown in Fig.~\ref{fig:LF}, together with fits of Eq.~(\ref{eq:H}) to the data. 
The rates are very small, close to the lower limit accessible to the technique, and the statistical error is large. Control experiments and precautions taken to minimize systematic errors are discussed in the Supplemental Material. 

Table~\ref{tab:H} gives values of $\tau_c$ and $B_\mathrm{loc}^\mathrm{\,rms}$ from the fits, together with one-$\sigma$ statistical uncertainties. 
\begin{table} [ht]
\begin{minipage}{0.48\textwidth}
\centering
\caption{\label{tab:H} Correlation times~$\tau_c$ and rms muon local fields~$B_\mathrm{loc}^\mathrm{\,rms}$ from muon spin relaxation rates in \ybco{y}.}
\begin{ruledtabular}
\begin{tabular}{cccc}
$y$ & Temperature (K) & $\tau_c$ (ns) & $B_\mathrm{loc}^\mathrm{\,rms}$ (mT) \\
\hline
6.72 & 80 & 5(2) & 0.92(19) \\
6.77 & 85 & 10(3) & 0.87(10) \\
6.83 & 93 & 25(10) & 0.37(6)
\end{tabular}
\end{ruledtabular}
\end{minipage}
\end{table}
The experimental values of $B_\mathrm{loc}^\mathrm{\,rms}$ differ from zero by 5--9$\sigma$ individually and ${\sim}10\sigma$ cumulatively; nonzero values are established at this level. Both $\tau_c$ and $B_\mathrm{loc}^\mathrm{\,rms}$ vary smoothly with~$y$. We have carried out dipolar lattice sum calculations for $B_\mathrm{loc}$, assuming candidate muon stopping sites from the literature~\cite{NiMi91, *WBGH91} and approximating the current loops as point dipoles~\cite{SSAM08}. These yield estimates~$B_\mathrm{loc} \approx 1$--1.5~mT for 0.1-$\mu_B$ loop-current magnetic moments, of the same order of magnitude as the observed values. The calculated values are not changed significantly for the ``criss-cross'' bilayer loop-current configurations recently reported by Mangin-Thro \textit{et al.}~\cite{M-TLSB17}.

It can be seen that $\tau_c$ falls in the middle of the range of experimental consistency discussed above. The observed increase of $\tau_c$ with increasing $y$ could be due to the approach to a quantum critical point as $T_\mathrm{mag} \to 0$. But fluctuations of the short-range IUC magnetic order may be associated with defects~\cite{Varma14}, in which case $\tau_c$ could depend on sample preparation and not be an intrinsic property. More work is required to elucidate the nature of the observed fluctuations.

\begin{figure*} [ht]
 \begin{center}
 \includegraphics[clip=,width=0.9\textwidth]{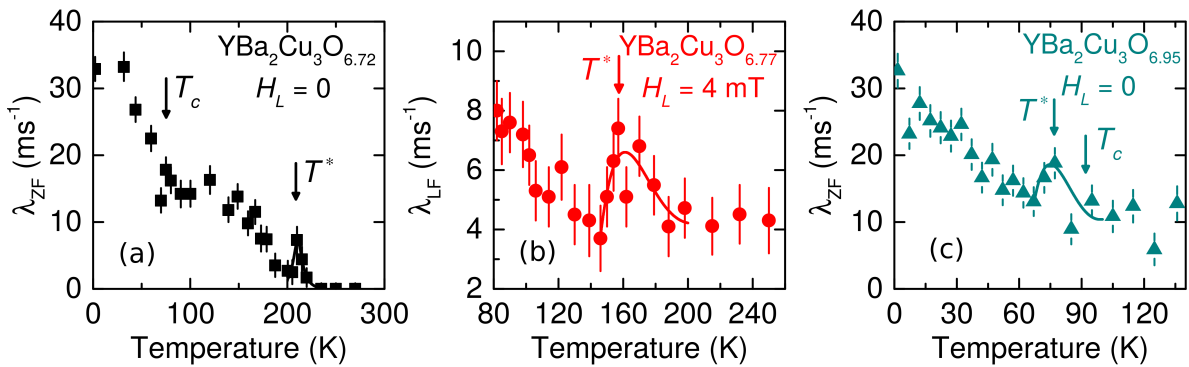}
\caption{(Color online) Temperature dependence of the dynamic muon relaxation rate $\lambda$ in YBa$_2$Cu$_3$O$_y$. (a)~$y = 6.72,$ longitudinal field~$H_L = 0$. (b)~$y = 6.77$,\ $\mu_0H_L = 4$~mT\@. (c)~$y = 6.95,\ H_L = 0$. The pseudogap onset temperature~$T^\ast$ is shown for each doping.}
 \label{fig:TDep}
 \end{center}
\end{figure*}

Exponential relaxation is observed in ZF, together with the expected Gaussian contribution due to random quasistatic dipolar fields from nuclear moments. Just above $T_c$ $\lambda_\mathrm{ZF}$ for $y = 6.72$ [Fig.~\ref{fig:TDep}(a)] is a factor of about 5 higher than $\lambda_\mathrm{LF}$ above $T_c$ [Fig.~\ref{fig:LF}(a)]. Some of this increase is due to a Lorentzian contribution to the distribution of static fields~\cite{Sonier01}, but some of it is doubtless due to dynamic relaxation; in ZF the two are hard to disentangle experimentally (see Supplemental Material). 

The temperature dependence of $\lambda_\mathrm{LF}$ in \ybco{6.77} with $\mu_0H_L = 4$~mT is shown in Fig.~\ref{fig:TDep}(b), and Figs.~\ref{fig:TDep}(a) and \ref{fig:TDep}(c) show the exponential relaxation rate~$\lambda_\mathrm{ZF}(T)$ measured in \ybco{6.72} and \ybco{6.95}, respectively. 
Maxima at $T_\mathrm{mag} \approx T^\ast$ and low-temperature increases are observed in all samples and fields, with statistical significance levels of 4--5$\sigma$ individually and ${\sim}8\sigma$ cumulatively (see Supplemental Material). A relaxation-rate maximum is often observed at second-order magnetic transitions~\cite{[{See, e.g., }]Dalmas97}, associated with critical slowing down of magnetic fluctuations near $T_\mathrm{mag}$. The low-temperature increase is unusual, however, and is discussed below in more detail.

\begin{figure*} [ht]
 \begin{center}
 \includegraphics[clip=,width=0.9\textwidth]{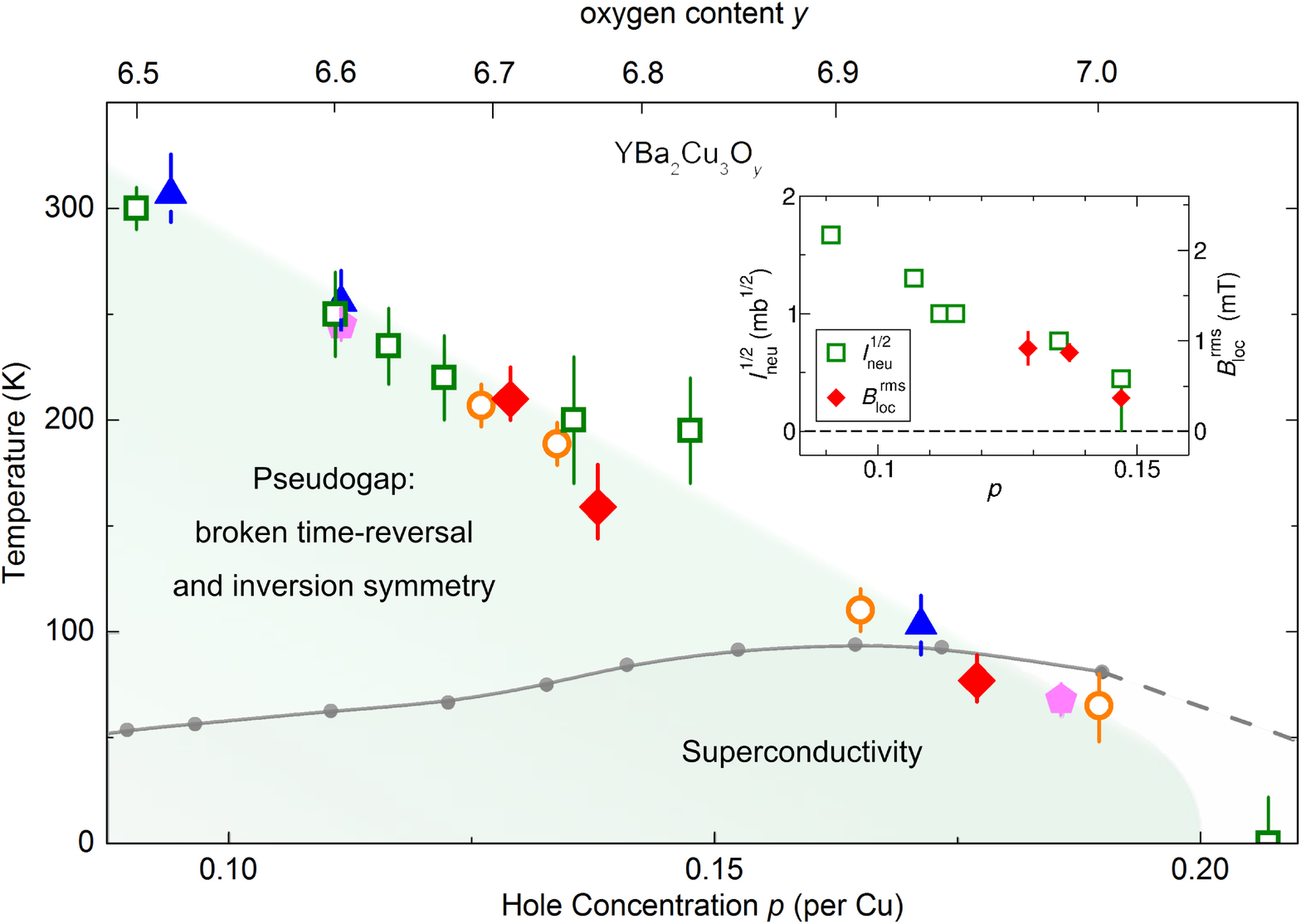}
\caption{(Color online) Phase diagram of \ybco{y}. $T_\mathrm{mag}$ (red diamonds) from the maximum in the temperate dependence of $\lambda$ (Fig.~\ref{fig:TDep}). Pseudogap temperatures $T^\ast$ determined from THz birefringence~\cite{Lubashevsky14} (blue triangles), resonant ultrasound~\cite{Shekhter13} (pink pentagons), polarized neutron scattering~\cite{Fauque06, *M-TSWB15} (green squares), and second harmonic generation~\cite{Zhao16} (orange circles). Grey points: superconducting transition temperatures. Inset: doping dependences of the square root~$I_\mathrm{neu}^{1/2}$ of the polarized neutron diffraction cross section~\cite{Li08} and the rms magnitude~$B_\mathrm{loc}^\mathrm{\,rms}$ of the fluctuating local field (Table~\ref{tab:H}).}
 \label{fig:phase}
\end{center}
\end{figure*}

In the phase diagram of Fig.~\ref{fig:phase}, $T_\mathrm{mag}$ from our $\mu$SR data is plotted versus the hole concentration $p$ (and the oxygen content $y$) along with $T_c$ and $T^\ast$ from other experiments. 
Values of $T_\mathrm{mag}$ are consistent with results of other experiments: polarized neutron scattering~\cite{Fauque06, *M-TSWB15}, THz birefringence~\cite{Lubashevsky14}, resonant ultrasound~\cite{Shekhter13}, and second harmonic generation~\cite{Zhao16}. ($T^\ast$ is also the temperature around which changes in transport properties~\cite{Nernst} and thermodynamic properties~\cite{Leridon, *Loram} from those of the strange metal phase begin to be observed.) The inset of Fig.~\ref{fig:phase} compares the doping dependence of the square root~$I_\mathrm{neu}^{1/2}$ of the polarized neutron diffraction cross section~\cite{Li08} with that of $B_\mathrm{loc}^\mathrm{\,rms}$. These quantities are both proportional to the order parameter for IUC magnetic order, and they follow the same trend.

Our observed temperature dependence of $\lambda_\mathrm{ZF}$ is consistent with earlier $\mu$SR measurements in two samples with $y = 6.67$ and $y = 6.95$~\cite{Sonier01}, except that in those experiments no maxima were seen near $T^\ast$. These may have been missed because the temperatures of measurement were too widely spaced; we also note that for $y = 6.67$ $T^\ast \approx 220$~K is in the region where relaxation due to muon hopping obscures the maximum (see Supplemental Material). 

Previous transverse-field (TF) $\mu$SR experiments in the pseudogap phase~\cite{[{See, e.g., }] LMCdMI13} observed exponential relaxation and ascribed it to static spatial inhomogeneity of superconducting fluctuations. Our observed LF-$\mu$SR rates (Fig.~\ref{fig:LF}) are an order of magnitude slower than the TF-$\mu$SR rates. This is consistent with the assumption that the latter are static, and precludes detecting the dynamic relaxation in TF-$\mu$SR. 

The observed increase of motionally narrowed muon spin relaxation with decreasing temperature below $T^\ast$ shown in Fig.~\ref{fig:TDep} cannot occur in a transition to a uniform ordered state. It is, however, consistent with low frequency fluctuations in domains of IUC magnetic order, of increasing magnitude with increasing order parameter~\cite{Varma14}. Scanning tunneling spectroscopy experiments~\cite{Davis} have found such domains in the pseudogap phase associated with defects, with linear dimensions of $\sim$20 unit cells; these provide a mechanism for the pseudogap in the one-particle spectra~\cite{Varma14}. Other experiments~\cite{Orenstein, *UBC} have observed mysterious anomalous low-frequency fluctuations in the pseudogap phase ascribed to extended defects. 

Wu et al.~\cite{WMKH15} summarize evidence for charge-density-wave (CDW) formation in \ybco{y}\ from NMR/NQR and other results. For $y = 0.672$ the onset temperature for short-range CDW (SRCDW) is 120-130 K, well below $T_\mathrm{mag} \approx 210$~K and the region where lambda increases with decreasing temperature [Fig.~\ref{fig:TDep}(a)]. For $y = 6.77$ the interpolated SRCDW onset from Wu et al. is at $\sim$100~K, again well below $T_\mathrm{mag} \approx 160$~K\@. Data for that sample were taken in LF, so that the relaxation is purely dynamic. For $y = 6.95$ ($T_\mathrm{mag} \approx 80$~K) the extrapolated SRCDW onset is at 20-30~K~\cite{WMKH15}. For all $y$ the onset of long-range CDW is at still lower temperatures, and is only observed in high applied fields. Furthermore, recent Nernst-coefficient measurements~\cite{C-CDLC17u} indicate that superconducting fluctuations persist only slightly above $T_c$, so that ``\dots the pseudogap phase is not a form of precursor superconductivity''. Thus there is no evidence for associating the increased low-temperature relaxation with either CDW formation or superconducting fluctuations.

The observed critical slowing down at $T_\mathrm{mag} \approx T^\ast$ (Fig.~\ref{fig:phase}) indicates that this temperature marks the onset of broken time-reversal symmetry. It is the only symmetry breaking observed to occur starting at $T^\ast$, and it is found in each of the four hole-doped cuprate families amenable so far to polarized neutron scattering experiments. The observed magnitude of the order parameter of about 0.1$\mu_B$ staggered moment per unit cell has a condensation energy $\sim$50~J/mol~\cite{VaZh15}, similar to the maximum superconducting energy in cuprates. These properties are all consistent with our results. 

Most importantly: our discovery of fluctuating magnetic fields provides an understanding of the absence of static magnetic fields due to IUC magnetic order in \ybco{y}. The expected fields are present but fluctuating. Although $\mu$SR is a point probe in real space, and thus is not directly sensitive to the spatial symmetry, our results are strong evidence for IUC order and its excitations, and establishes them as important for understanding the unusual behavior of cuprates.

{\bf Acknowledgements}\quad We are grateful to C.~M. Var\-ma for proposing these experiments and for numerous discussions. We thank the support teams at TRIUMF and ISIS for their help during the experiments, and J.~H. Brewer 
for discussions and correspondence. The research performed in this study was partially supported by the National Natural Science Foundation of China No.~11474060, STCSM of China Grant No.~15XD1500200, MOST of China No.~2016YFA0300403. Research at U.C. Riverside was supported by the UC Riverside Academic Senate. Work at CSULA was funded by NSF/DMR/PREM-1523588. Research at CSU-Fresno was supported by NSF DMR-1506677.

%


\newpage

\setcounter{equation}{0} \setcounter{figure}{0} \setcounter{table}{0}

\noindent \textbf{SUPPLEMENTAL MATERIAL}

\paragraph{Sample growth and characterization.} High quality single crystals of \ybco{y}~were grown by the top-seeded solution-growth (TSSG) polythermal method using 3BaO-5CuO solvent~\cite{Xiang16S}. A \ybco{y} single crystal with an $ab$ plane area of $10{\times}10~\mathrm{mm}^{2}$ and $c$-axis length of 8~mm was synthesized with a cooling rate of 0.5~K per hour in air. The crystal was then cut into small pellets with thickness~0.55~mm and lateral dimensions~2~mm$\times$2~mm. Single crystals with optimal $T_c = 91$~K were achieved by annealing at 400$^{\circ}$ for 180 h in flowing oxygen. A range of oxygen concentrations of \ybco{y} was achieved by post-annealing in flowing oxygen at different temperatures as described in Ref.~\onlinecite{Gao06S}, resulting in superconducting transition temperatures between 72 K and 88 K\@. 

\begin{figure} [ht]
\includegraphics[clip=,width=3.2in]{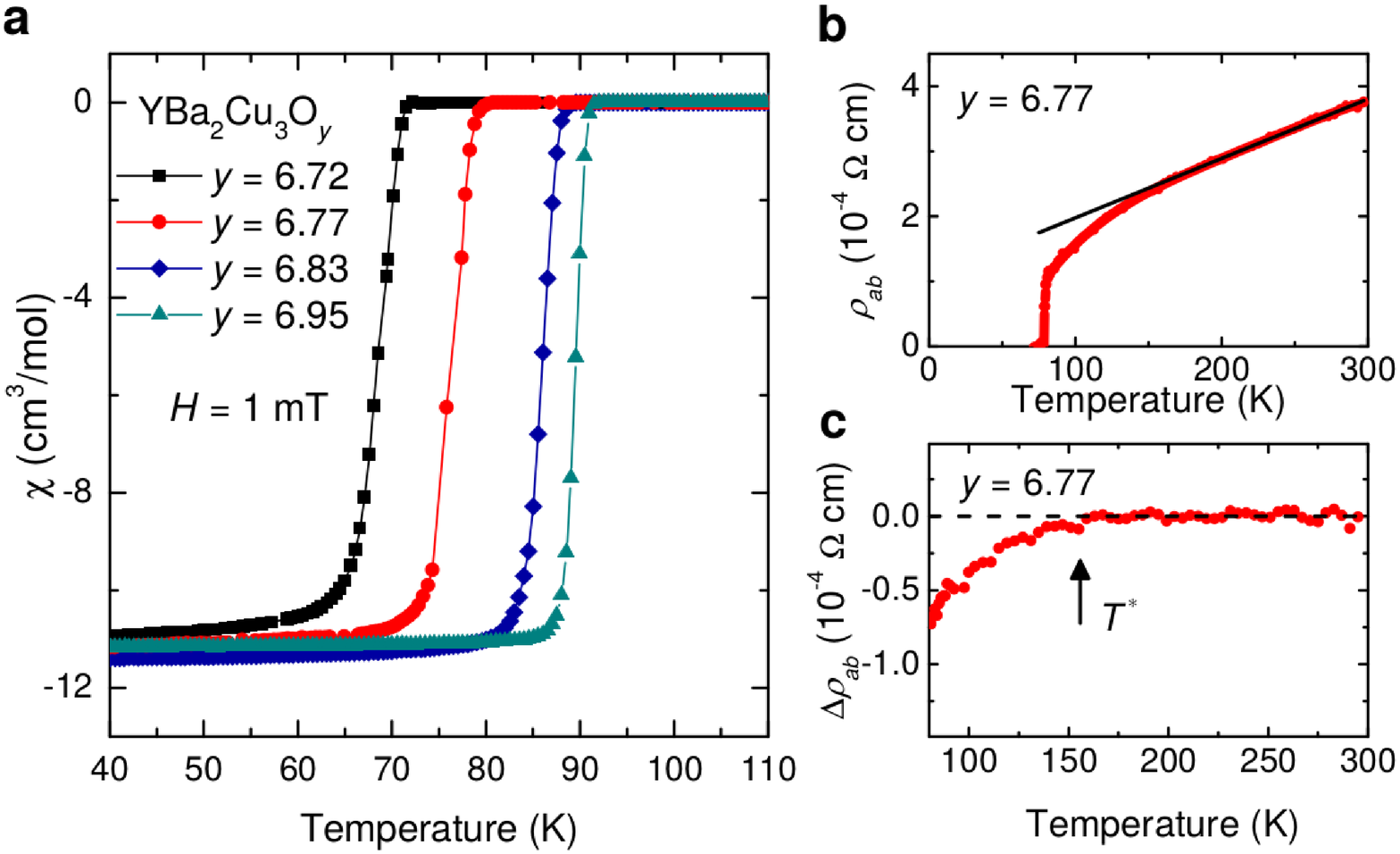}
\caption{\label{fig:squid2} (color online) Characterization data from \ybco{y} single crystals. (a)~Magnetization of \ybco{y}, $y= 6.72$, 6.77, 6.83, and 6.95, showing sharp superconducting transitions. (b)~Temperature dependence of electrical resistivity $\rho_{\rm ab}$ for $y= 6.77$ showing a deviation from linearity. (c)~Temperature dependence of the deviation~$\Delta\rho_{\rm ab}$ from linear resistivity for $y= 6.77$. The deviation sets in below $T^{\ast} = 156$~K, which is consistent with reported results~\cite{Ando04S}.}
\end{figure}

Figure~\ref{fig:squid2} 
shows the temperature dependences of the magnetization and the resistivity in our samples. The data indicate that the superconducting transitions are sharp. The values of $T^\ast$ from the departure of the resistivity from linearity are in good agreement with previous results~\cite{Ando04S}.

\paragraph{Muon spin relaxation experiments.} Our $\mu$SR experiments were performed using the LAMPF spectrometer at the M20 beam line of TRIUMF, Vancouver, Canada, and the MUSR and EMU spectrometers at ISIS, Rutherford Appleton Laboratory, Chilton, U.K\@. At both facilities muons were implanted into the sample with their initial spin polarization~$\mathbf{P}_\mu$ perpendicular to the $ab$ plane. 

\begin{figure} [ht]
\includegraphics[clip=,width=3.2in]{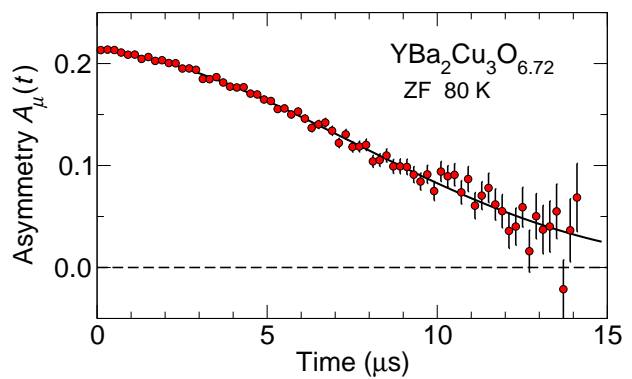}
\caption{\label{fig:ZF} (Color online) Zero-field relaxation of the muon asymmetry~$A_\mu(t)$ in \ybco{6.72}. Data taken at 80~K\@. Curve: fit of Eq.~(\ref{eq:damp}) to the data. The need for exponential damping is clear from the nonzero initial slope of $A_\mu(t)$; this would vanish for an undamped Gaussian.}
\end{figure}

Static relaxation in ZF-$\mu$SR reflects the distribution of random static local fields. A Gaussian distribution of field components, which is a good approximation to the dipole-field distribution expected from randomly-oriented (quasi)static nuclear magnetic moments, gives rise to the so-called static Gaussian ZF Kubo-Toyabe relaxation function~\cite{KT67S,Hayano79S}
\begin{equation}
A_\mu^\mathrm{KT}(t) = A_\mu(0) \left[\frac{1}{3}+\frac{2}{3}(1-\sigma ^2 t^2)\exp(-{\textstyle\frac{1}{2}}\sigma^2t^2)\right]
\label{eq:AKT}
\end{equation} 
for the asymmetry~$A_\mu(t)$, where $\sigma/\gamma_\mu$ is the rms width of the distribution. ZF relaxation due to an additional dynamic fluctuating field is often modeled by exponential damping of $A_\mu^\mathrm{KT}(t)$:
\begin{equation} 
A_\mu(t) = e^{-\lambda t}A_\mu^\mathrm{KT}(t)\,.
\label{eq:damp}
\end{equation} 
Figure~\ref{fig:ZF} shows typical ZF asymmetry relaxation. The fit of Eq.~(\ref{eq:damp}) to the data yields $\sigma/\gamma_\mu  = 0.099(2)$~mT, a typical nuclear dipolar field value.

In ZF it is difficult to discriminate between static and dynamic exponential relaxation~\cite{Maisuradze10S}, but for $H_L \gg \sigma/\gamma_\mu$ the static local fields are `decoupled'~\cite{Hayano79S} and the observed relaxation is purely dynamic. Our LF experiments were carried out for $H_L \gtrsim 2~\text{mT} \approx 20\sigma/\gamma_\mu$.

\begin{figure}
\includegraphics[clip=,width=\columnwidth]{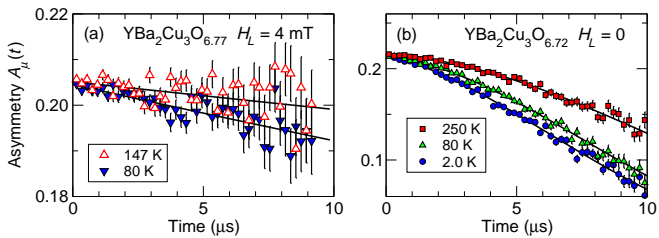}
\caption{\label{fig:ZFLFasy} (Color online) Time evolution of the positron count-rate asymmetry~$A_\mu$ at various temperatures and fields in single-crystal \ybco{y}. (a)~\ybco{6.77}, longitudinal magnetic field~$H_L = 4$~mT. Curves: exponential fits. (b)~\ybco{6.72}, zero field. Curves: fits of Eq.~(\ref{eq:damp}) to the data.}
\end{figure}

Appropriate functional forms were least-squares fit to the asymmetry data using the \textsc{musrfit} $\mu$SR analysis program~\cite{Suter12S}. LF asymmetry data were fit by a simple exponential $A_\mu(t) = A_\mu(0) \exp(-\lambda_\mathrm{LF} t)$. An example is given in Fig.~\ref{fig:ZFLFasy}(a), 
which shows the asymmetry in \ybco{6.77} for two temperatures at LF field~$H_L = 4$~mT\@. The LF rates are very low and are near the resolution limit of the technique. The difference between the rates is small but resolved. In ZF the relaxation is dominated by the Gaussian static nuclear dipolar field distribution [Fig.~\ref{fig:ZFLFasy}(b)]. The Gaussian rate decreases at high temperatures due to the onset of muon hopping~\cite{Sonier02bS}. The exponential contribution to the relaxation, visible as a nonzero initial slope, also decreases with increasing temperature. 

\paragraph{Statistical significance.} The inverse relative standard deviations (IRSD) (the $N$ in ``$N\sigma$'') for $B_\mathrm{loc}^\mathrm{rms}$ (article, Table~I) and for the maxima in the temperature dependences at $T_\mathrm{mag}$ (article, Fig.~2), are shown in Table~\ref{tab:sigma}\@.
\begin{table} [ht]
\centering
\begin{minipage}{0.48\textwidth}
\caption{\label{tab:sigma} Inverse relative standard deviations (IRSD) of$B_\mathrm{loc}^\mathrm{rms}$ and maxima amplitudes near $T_\mathrm{mag}$ from muon spin relaxation rates in \ybco{y}.}
\begin{ruledtabular}
\begin{tabular}{cccc}
 & $B_\mathrm{loc}^\mathrm{rms}$ & \multicolumn{2}{c}{maxima (Fig.~2)}\\
$y$ & IRSD & No.~of points & IRSD \\
\hline
6.72 & 4.8 & 7 & 4.5 \\
6.77 & 8.7 & 15 & 5.2 \\
6.83 & 6.2 & &\\ 
6.95 & & 6 & 3.9 \\
cumulative & 11.7 & & 7.9
\end{tabular}
\end{ruledtabular}
\end{minipage}
\end{table}

\noindent $N$ for $B_\mathrm{loc}^\mathrm{rms}$ is simply its value divided by its standard deviation. For the maxima at $T_\mathrm{mag}$, baseline points were chosen above and below each maximum, and baseline values at intermediate points were estimated by linear interpolation. The IRSD of each point is then its amplitude relative to the baseline divided by its standard deviation, and the IRSD of the maximum is the square root of the sum of squares of the IRSDs of the points. The sign of the amplitude is included in this sum to account for negative contributions. In both cases the cumulative IRSD is the square root of the sum of squares of the individual sample IRSDs. Using the usual criterion~$N \gtrsim 5$ for statistical significance, it can be seen that some individual IRSDs are marginal but that the cumulative values are quite satisfactory.

\paragraph{Control experiments.} The muon radioactive lifetime~$\tau_\mu = 2.197~\mu$s limits the minimum measurable muon spin relaxation rate to $\gtrsim 10^3~\mathrm{s}^{-1}$ for ${\sim}10^8$ positron counts total. Our measured rates are of this order, so that long data runs are needed. In addition, considerable care must be taken to avoid systematic errors. We have carried out control experiments on silver samples at both TRIUMF and ISIS as measures of such errors. In absolutely pure silver the only appreciable sources of local magnetic fields are $^{107}$Ag and $^{109}$Ag nuclear moments; these are quite small, making silver an ideal material for control experiments (and $\mu$SR sample holders). Bueno \textit{et al.}~\cite{Bueno11S} have reported a careful LF-$\mu$SR study at TRIUMF of high-purity silver at room temperature in a LF of 2~T\@. Systematic errors from muons that stop in the trigger scintillator, errors in fit parameters other than the relaxation rate, and time structure in the putative uncorrelated background were considered. They report an exponential relaxation rate~$\lambda_\mathrm{Ag} = 1.0 \pm 0.2 \text{ (statistical)} \pm 0.2 \text{ (systematic) ms}^{-1}$.

Our control experiments were performed on two silver sheet samples, designated Ag1 and Ag2, both of 6N purity from Alfa-Aesar. Uncorrelated background count rates, trigger scintillator thickness, and other conditions at TRIUMF were comparable to those of Ref.~\onlinecite{Bueno11S}. Sample~Ag1, which was measured at both TRIUMF and ISIS, is a 10~mm$\times$10~mm square, whereas sample~Ag2, which was measured only at ISIS, is large enough to stop the entire ISIS beam. Ag1 and the \ybco{y} samples were mounted on Ag2 at ISIS, which served as a cold finger and contributed a fraction~$f_\mathrm{bkgd}$ to the observed muon polarization. This fraction is difficult to determine when all the relaxation rates are low; for Ag1 we estimate $f_\mathrm{bkgd} \approx 0.3$ from its dimensions and those of the cold finger.

\begin{figure} [ht]
\includegraphics[clip=,width=3.2in]{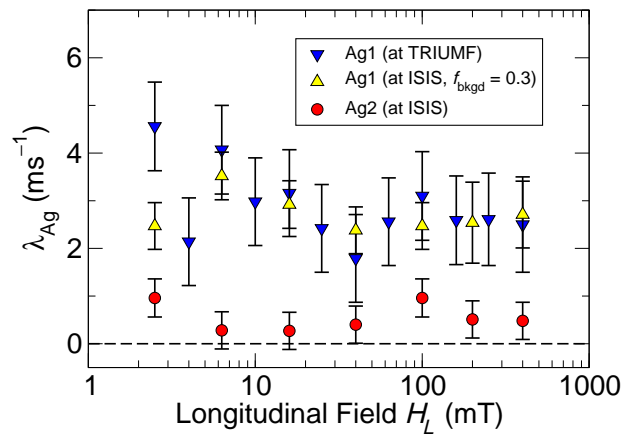}
\caption{\label{fig:Ag-RT} (Color online) longitudinal-field muon spin relaxation rates in silver samples. Data taken at room temperature, using the LAMPF spectrometer at TRIUMF and the EMU spectrometer at ISIS.}
\end{figure}

The field dependences of $\lambda_\mathrm{Ag}$ for both samples at room temperature are shown in Fig.~\ref{fig:Ag-RT}.
There is no statistically significant field dependence. For Ag1, data taken at TRIUMF and ISIS yield consistent rates (averages $2.9 \pm 0.8~\text{ms}^{-1}$ and $2.7 \pm 0.4~\text{ms}^{-1}$, respectively; uncertainties are statistical only). The rates for Ag2 are significantly lower than for Ag1. The average rate for Ag2, $0.5 \pm 0.3~\text{ms}^{-1}$, is consistent with with the result of Ref.~\onlinecite{Bueno11S}, in which it is noted that at this level small concentrations of impurities and/or defects can dominate the relaxation and lead to differences between nominally pure samples.

From the consistent results for the two spectrometers, we (1)~attribute the difference between rates in Ag1 and Ag2 to a higher concentration of defects or impurities in the former, and (2)~conclude that uncontrolled sources of systematic error in our experiments are not a major factor at either facility.

We also carried out an experiment on the LAMPF spectrometer at TRIUMF with no sample, to determine the fraction of the muons that are scattered by the muon counter, miss the sample, stop elsewhere in the spectrometer, and generate spurious positron counts. The fraction is small ($\sim$2\%) but not negligible, and may account for a portion of the observed background rate; the signal was too weak to allow a reliable determination of its relaxation rate.

\paragraph{Muon hopping, superconductivity.} We have limited the range of hole doping of our samples to the region for which $T^\ast\lesssim 200$~K, because above this temperature thermally-activated hopping of muons~\cite{Hayano79S} becomes important~\cite{Sonier01S}; this causes unwanted dynamic relaxation by nuclear dipolar fields. For $T_\mathrm{mag}$ less than the superconducting transition temperature we consider only ZF data, since field cooling into the superconducting region, even at constant field, might lead to complications from flux pinning and dynamics.

\paragraph{High temperature relaxation.} In all samples a constant rate of 1--3~ms$^{-1}$ is seen at high temperatures. Its origin is not yet understood, but is unlikely to be systematic error since the control experiments on silver samples discussed above show no sign of such an effect. Korringa relaxation due to spin fluctuations in band states~\cite{Slichter96S} can definitely be ruled out. The Korringa rate~$\lambda_K$ is given by
\begin{equation} \label {eq:Korr}
\lambda_K \approx \hbar \gamma_\mu^2 \langle B_\mathrm{loc}^{\,2}\rangle k_BT/E_F^2 \,,
\end{equation}
where $E_F$ is the Fermi energy. Assuming $E_F \sim 1$~eV, Eq.~(\ref{eq:Korr}) requires $B_\mathrm{loc}^\mathrm{\,rms} \sim 10$~T to produce the observed rates~$\sim 10^3~\text{s}^{-1}$ at $T \sim 100$~K\@. This is a much larger muon local field than any known mechanism could provide.

\end{document}